# NON-STATIONARY THERMOPHYSICAL CHARACTERIZATION OF EXFOLIATED GRAPHITE WITH CARBON NANOTUBES COMPOSITES

N. V. Morozovsky[1], Yu. M. Barabash[1], Yu. V. Grebelna[2], M. T. Kartel[3], Yu. I. Sementsov[3], G. I. Dovbeshko[1]

[1]Institute of Physics, NAS of Ukraine, 46 Nauky Ave., 03028 Kyiv, Ukraine
[2]TM Spetsmash ltd., 02094, Viskozna str., 5, Kyiv, Ukraine
[3]O. Chuiko Institute of Surface Chemistry, NAS of Ukraine, 03164, Kyiv, Ukraine
e-mail: gd@iop.kiev.ua

The sheet samples of thermally exfoliated graphite (TEG) - carbon nanotubes (CNT) composites (TEG-CNT-cs) were obtained by persulphate oxidation using chemical (CO) and electrochemical (anode) oxidation (ECAO). Electron microscopy reveals multi-layered structures of few-layer graphene nanosheets with folded and tubular-like fragments. The effective thermal diffusivity values were estimated by non-stationary photo-pyroelectric thermophysical characterization using the heat pulse and thermo-wave modulation methods. Comparison with other carbon (C-) based thermal management materials shows that TEG-CNT-cs exhibit thermal diffusivity, effusivity and conductivity comparable with those of actual C-polymer- and C-C-composites. For TEG-CNT-cs, evaluated values of phonon mean free path (MFP) and relaxation time (RT) are in the ranges estimated for defective graphene. The values of diffusivity and effusivity, MFP and RT are lower for denser TEG-CNT-cs obtained by ECAO and are higher for less dense TEG-CNT-cs obtained by CO. The obtained diffusivity and effusivity values designate TEG-CNT-cs as suitable thermal management materials.

## 1. INTRODUCTION

Modern carbonaceous (C-) materials, including single wall (SW) and multiwall (MW) carbon nanotubes (CNT) [1 - 8], CNT-based structures (mats [9 - 11], bundles [11, 1], yarns [13, 14], sheets [15, 16] and foams [17], as well as graphene [18 - 21] and exfoliated graphene [22, 23], exfoliated [24 - 26] and expanded [10, 27, 28] graphite, carbon-carbon [10] and carbon-polymer [29 - 31] composites and hybrids [21, 31] are promising to application in various science intensive technological areas [2 - 8, 20 -23, 25, 27, 29, 31 - 36].

Materials based on graphene and CNTs are also promising for use in biosensors [37] and AFM biotechnologies [38], respectively.

The large spectrum of C-based materials potentialities in industries [32] covers the applications in automobile [5, 16], aerospace [29, 33, 35] and spacecraft [2, 36], in particular for electromechanical actuators [34], for fuel lines electrical discharging [5], for heavy detail replacement [16], for spacecraft and aircraft high strength light weight body parts [2], for space micropropulsion thrusters [36] and for thermal management in electronic devices packages [10, 39 - 43] as thermal interface and/or heat sink materials [43 - 54].

C-C-materials based on thermally exfoliated [25, 26] and expanded [28, 55] graphite have been used as gland seals of friction units and flange gaskets of valves in power pipelines, centrifugal and vortex pumps, heat exchange vessels and as refractory sealing, fire extinguishing and flame-retardant agents in other equipment of thermal and nuclear power plants and petrochemical enterprises [25, 26, 28, 55].

At a matter of fact, namely CNT- and graphene- based composites, rather than individual CNTs and single-layer graphene, are used for large-scale technical applications, such as highly efficient thermal interface materials [44 - 47] and in thermal management [10, 29, 35, 40 - 43, 49, 50 - 54].

Under operation of any C-material in the thermal management as an IR-absorber [49 - 53] (IR-to-heat energy converting), as a thermal interface material [44 - 47] or as a heat sink [31, 48], the most important are its heat transfer and heat accumulation abilities.

Under stationary conditions, these qualities are determined by two parameters: thermal conductivity, $k_T$, and mass specific heat capacity, $c_p$.

In their turn, $k_T$ and $c_p$ determine other two parameters: thermal diffusivity, $\alpha_T = k_T/C_\rho$, and thermal effusivity, $b_T = (k_T C_\rho)^{1/2} = C_\rho a_T^{1/2}$, where $C_\rho = c_p \rho$ is the volume specific heat capacity, $\rho$ is the density of material [56].

The knowledge of $a_T$ and $b_T$ is especially important in the case of operation at the non-stationary conditions of thermal management, when quick heat exchange (high $a_T$) and effective heat sink (high $b_T$) must be realized.

The thermo-physical parameters have been studied for a variety of C-materials [9Marcon13], including SWCNT [11, 57, 58], MWCNT [59, 60], SWCNT-MWCNT [61] composites, low density (0.38 g/cm$^3$) CNT-sheets [16], MWCNT-films [66], graphene [42, 43], graphene laminate films [31], expanded graphite-CNT [10] and graphene-graphene [45] nanocomposites, graphene-CNT and graphene-graphite composites [48], as well as hybrid graphene-metal nano-micro-composites [46].

Brief reviews of the results of theoretical estimations and experimental measurements of thermal diffusivity $a_T$ and thermal conductivity $k_T$ of CNTs, graphene and a number of materials based on them evidences high spread of obtained $a_T$ and $k_T$ values.

Thus, for various C-materials, the obtained $a_T$ values are ranged from ~$10^{-5}$ to ~$10^{-3}$ m$^2$/s [16, 52, 60 - 65] and the obtained $k_T$ values are ranged from ~1 to ~$10^3$ W/m·K [9, 11, 20, 26, 30, 31, 47, 48, 66] depending on the microstructure, morphology and direction of measurements (see Section 3.1 for details).

Therefore, any information about the thermophysical properties of each specific C-C-composite material is desirable.

The lack of data on the thermophysical parameters of such C-C-materials as nanocomposites based on thermally exfoliated graphite (TEG) and CNT (TEG-CNT-cs) determines the advisability of their study.

In this paper, are presented the results of non-stationary thermo-physical characterization of sheet shaped samples of TEG-CNT-cs synthesised by persulphate oxidation method.

## 2. EXPERIMENTAL

### 2.1. CNT preparation

The multiwall CNTs were synthesized by catalytic pyrolysis using catalytic chemical vapor deposition (CVD) technology [28, 67]. The complex oxide catalysts with a metal ratio (Al:Fe:Mo = 2:1:0,21) obtained by the aerosol method via decomposition of an aqueous mixture of iron citrate and aluminum and molybdenum formates were used. As the source of carbon, the propylene obtained by dehydration of propanol was used. Under optimal conditions, the degree of propylene to carbon conversion was about 80-97%. A pyrogenic silicon (aerosil) grade A 300 was added to the catalyst to prevent the agglomeration of CNTs in the process of synthesis. CNTs with the diameter from 2 to 40 nm were obtained using a 24 dm$^3$ rotating reactor for homogeneous mixing of the catalyst beds. TEM images of prepared CNTs are shown in the **Figure 1a**.

### 2.2. TEG preparation

The samples of thermally exfoliated graphite (TEG) were synthesised by persulphate oxidation method [28].

Strong acids [68] and oxidants [69] are known as the effective deagglomerating agents for MWCNT. It is reasonably to unite the processes of deagglomeration of CNT and intercalation of natural graphite. Such a procedure was performed by two ways.

The 1-st one, electrochemical (anode) oxidation (ECAO), consisted in the follows.

CNT in the amount of 1% of the weight of graphite were "dispersed" in 55% sulfuric acid, $H_2SO_4$, solution, and the natural graphite powder in the proportion of 1 kg of graphite to 1 $dm^3$ of $H_2SO_4$ was added. Then the mixture was processed by the amount of electricity of 90 A·h/kg.

The 2-nd one, chemical oxidation (CO), consisted in the follows. CNT in the amount of 2% of the weight of graphite was dispersed by mixing in a solution of potassium bichromate, $K_2Cr_2O_7$, in sulphuric acid with the ratio of 100 $cm^3$ of $H_2SO_4$ to 10 g of $K_2Cr_2O_7$. Then natural graphite powder in the amount of 50 g in 100 $cm^3$ of this solution was added to the CNT suspension.

In the both ECAO and CO ways, after the graphite oxidation to the first (blue) stage, it was hydrolyzed, washed to neutral pH, dried and heat treated at the temperature of ~ 1000 °C in an industrial horizontal gas furnace.

The sheets of TEG-CNT-cs were manufactured by rolling of the powder with horizontal rolls. Synthesis and rolling conditions were previously described [70 - 72] and implemented by TM Spetsmash Ltd. (Kyiv, Ukraine) [55].

The obtained EG and TEG-CNT-cs samples were characterized by transmission electron microscopy (TEM) [55, 73] using an electron microscope of EMV-100AK type at accelerating voltage up to 100 kV, which gave resolution up to 0.7 nm. The examples of the TEM images of TEG flakes and TEG-CNT-composite microstructure are shown in **Figure 1b,c,d**.

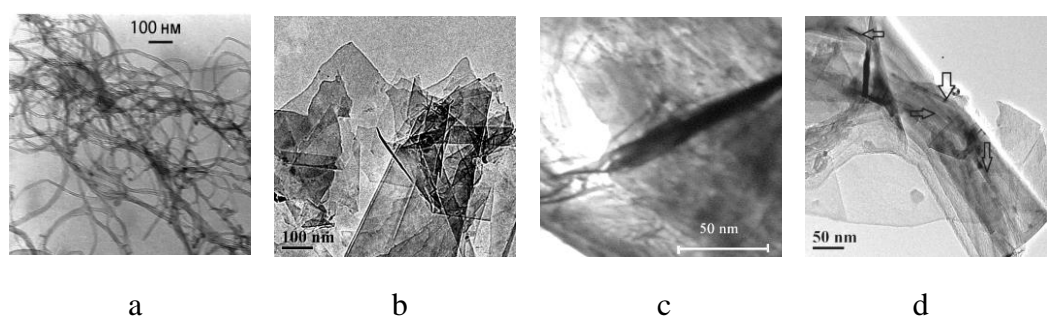

     a                           b                          c                          d

**Figure 1.** TEM images of CNTs (a), fragments of TEG flakes (b, c) and TEG-CNT composite (d) (adopted from [73]).

It can be seen (**Fig. 1b,c**) that in the structure of the studied TEG-CNT-cs there are the wrinkles and rough edges as well as the folded formations, possibly of a cylindrical shape . The sizes of the distinguishable nanostructure fragments range from 1 to 100 nm.

The picture shown in **Fig. 1c** is quite unique. The successful projection angle of the image reveals possible rolling up of the graphene flake into the nanoscroll. At that, the outer diameter of the cross-section of the nanoscroll can be estimated as ≈ 2 nm for its thin part, and as ≈20 nm for its thick part.

At the meso-scale, TEM reveals the aggregates of partially overlapped single layer graphene flakes (**Fig. 1b**), the folded, wrinkled and tubular-type structure fragments of few-layer graphene (**Fig. 1c**) and CNT "embedded" in the body of the EG scroll (**Fig. 1d**, indicated by arrows).

It should be noted the resemblance of different microstructure elements in TEM images shown for TEG-CNT-cs samples in **Figure 1b,c,d** and those obtained for graphene nanosheets - CNT samples presented in **Figure 1d,e,f** in the Ref. [48].

Thus, TEG-CNT-cs contain substructures within the TEG and, therefore, can be considered as a combination of two-dimensional forms of carbon (single-, bi- and few- layer graphene nanosheets) and their quasi-three-dimensional derivatives (folded and wrinkled graphene flakes), as well as three-dimensional derivatives (graphene scrolls, uncompletely exfoliated EG and CNTs).

**Table 1** contains the density values calculated by sample mass and dimensions, the graphite (G-) fraction, air fraction (porosity) and G/air fraction ratio, as well as the fraction of a dense-packed TEG in TEG-CNT samples calculated by dividing density of TEG-CNT-cs samples with the averaged density of graphite (≈ 2 g/cm$^3$) and packed TEG (≈ 1 g/cm$^3$) (see Tables 2.3, 2.9 in the Ref. [74].

**Table 1.** CNT percent, samples dimensions, mass and density, graphite (G) fraction, air fraction, G/air-fraction ratio, and the TEG fraction of a dense-packed TEG in TEG-CNT-cs samples of different preparation ways.

| Parameter/ Composite and Preparation | % CNT/ TEG | Thick-ness, μm | Dimen-sions, mm | Mass, mg | Density, mg/mm$^3$ (g/cm$^3$) | G-frac-tion, | Air-frac-tion | G/Air ratio | TEG-frac-tion |
|---|---|---|---|---|---|---|---|---|---|
| 1-E.Chem-1CNT | 1 | 250 | 7x16 | 10 | 0.357 | 0.18 | 0.82 | 0.22 | 0.36 |
| 2-Chem-CNT | 2 | 300 | 6x17 | 7.5 | 0.245 | 0,123 | 0.88 | 0.14 | 0,24 |
| 3-Chem-CNT | 2 | 450 | 7x17 | 12.5 | 0,233 | 0.115 | 0.885 | 0.13 | 0.23 |
| 4-Chem-sem | 2 | 500 | 6x17 | 12.0 | 0.235 | 0.118 | 0.88 | 0.14 | 0.24 |
| 5-Chem-China | 2 | 450 | 6x16 | 10 | 0,231 | 0.115 | 0.885 | 0.13 | 0.23 |
| 6-Elechem-mikle | 1 | 300 | 10x14 | 15 | 0,357 | 0.18 | 0.82 | 0.22 | 0.36 |

Notes:

Graphite (G) -fraction = $V_{Gr}/V = (m_G/\rho_G)/V \approx /m_G \approx m/ \approx (m/\rho_G)/V = (m/V)/\rho_G = \rho/\rho_G$ (Solid fraction);

Air-fraction = $V_{Air}/V = (V - V_G)/V = 1 - V_G/V = 1 - \rho/\rho_G = 1 - $ G-fraction (Porosity);

G-fraction/Air-fraction ratio = $V_G/(V - V_G) = (V_G/V)/[(V - V_G)/V] = 1/(\rho_G/\rho - 1)$ (Solid/Void fraction ratio);

TEG-fraction= $V_{TEG}/V = (m_{TEG}/\rho_{TEG})/V \approx \{m_{TEG} \approx m\} \approx (m/\rho_{TEG})/V = (m/V)/\rho_{TEG} = \rho/\rho_{TEG}$.

The decreased density of TEG ($\approx 1$ g/cm$^3$ [74]) in the comparison with the non exfoliated graphite ($\approx 2$ g/cm$^3$ [74]) can be associated, at least partially, with the increase of interlayer distance during oxidation process. Indeed, near twice increase of interlayer distance in the graphene oxide as a result of the few hours oxidation of graphite is well established [37].

The significant decrease in the density of TEG-CNT-cs as compared to TEG is primarily associated with the high degree of porosity, which determines the low fractions of graphite and TEG in TEG-CNT-cs (see **Table 1**).

The spread of the density values of porous TEG-CNT-cs in the limits from 0.23 g/cm$^3$ to 0.36 g/cm$^3$ corresponds to different graphite content (G-fraction) and so TEG content (TEG-fraction), as well as different content of air pockets (Air-fraction) defined the porosity degree in TEG-CNT-cs samples prepared by different ECAO and CO ways (see **Table 1**).

A similar density spread is typical, e. g., for graphene laminate layers (from 1.0 g/cm$^3$ to 1.9 g/cm$^3$) [31], where it depends on the average size and alignment of graphene flakes, and for vitreous carbon foams (from 0.025 g/cm$^3$ to 0.33 g/cm$^3$) with corresponding porosity (from 0.832 to 0.987) at a C-skeleton density 1.98 g/cm$^3$ [17].

The density of ECAO samples with 1 % of CNTs (samples 1 and 6 in the **Table 1**) is approximately 1.5 times higher than that of CO samples with 2 % of CNTs (samples 2 – 5 in the **Table 1**). This can be a consequence of not only the different porosity degree, but the different exfoliation level, which determines the spacing between locally separated grapheme platelets and other structure components of TEG-CNT-cs prepared by different ECAO and CO ways.

It is worth noting that TEG-CNT-cs, being flexible, have a density of only (20 - 30) % of the density of industrial flexible graphite ($\approx 1.1$ g/cm$^3$) (see [26], [28] and Ref. [53] ibid) and are no less thermally stable [74] than graphene/polymer composites [29, 30].

**2.3. Dynamic thermophysical characterization**

The both used in the present study photo-thermal methods, namely heat pulse method (HPM) [75, 76] and thermo-wave modulation method (TWM) [77, 78] belong to non-destructive non-contact photothermal radiometry techniques [79 - 82]. The HPMs are the followers of the flash method proposed by Parker et al. [83]. The TWMs are the followers of Ångström's method [84, 85] modified by Kogure and Hiki [86] and by Das et al. [87, 88].

In both HPM and TWM methods, the front surface of the sample is irradiated with a modulated "primary" radiation flux. In the HPM this modulation is pulse, while in the TWM it is sinusoidal. Optically absorbed part of this "primary" flux is transformed to the heat energy, which produces the temperature rise giving the start to temperature waves propagation into the sample. Certain part of the heat flow, being not scattered, gives temperature rise on the opposite rear surface of the sample, where it is transformed to "secondary" infrared (IR) radiation flux. This "secondary" IR radiation flux, which is delayed relatively to incident "primary" one due to finite speed of the heat diffusion process, is monitored by IR detector. If as IR detector the pyroelectric sensor is used [75, 76, 77, 78, 89], it is not necessary to know the values of the incident and absorbed energies, the absorptivity of the front face and the emissivity of the back face of the sample, as well as the temperature-time rise curve and pyroelectric parameters of the sensor [76, 89].

The block diagrams of the measurement systems HPM and TWM are presented in the **Figure 2a** and **Figure 3a**, respectively**.**

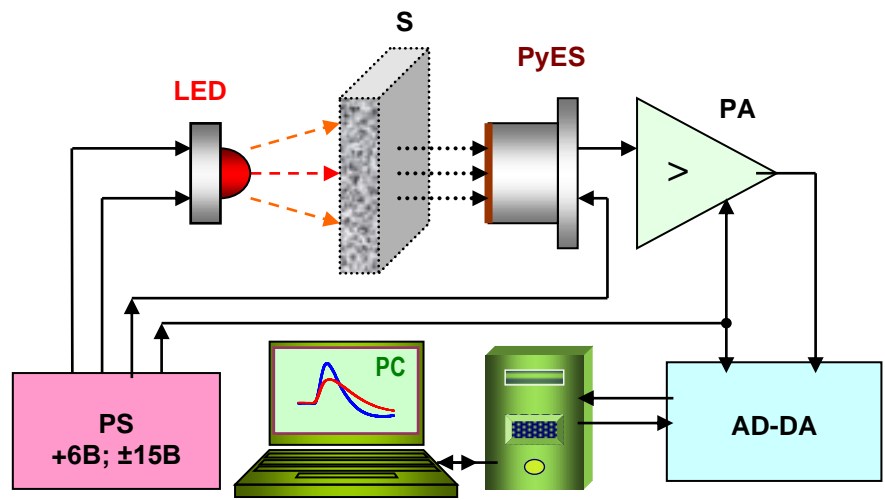

(**a**)

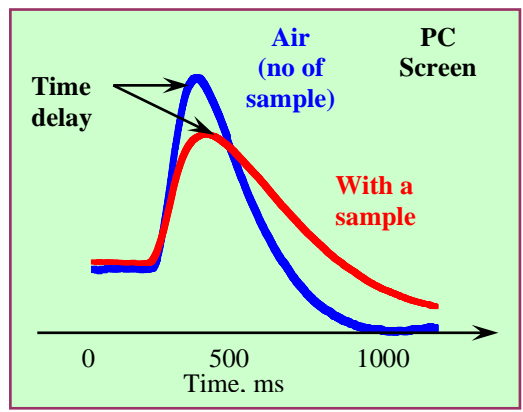

(**b**)

**Figure 2.** (a) - Block diagram of the heat pulse method.

LED – is the light emitting diode (GREE-XML; 3V, 1 W), S – is the sample under test, PyES – is the pyroelectric sensor (IRA-E700ST0), PA – is the preliminary amplifier (Ku=10000); AD-DA – is the analog-digital – digital-analog converter, (±10B, 12 resol., 2 charnels), PS – is the power supply (+6B; ±15B), PC – is the personal computer.

(b) - Illustration of time delay between pyroelectric responses on "primary" and "secondary" heat fluxes.

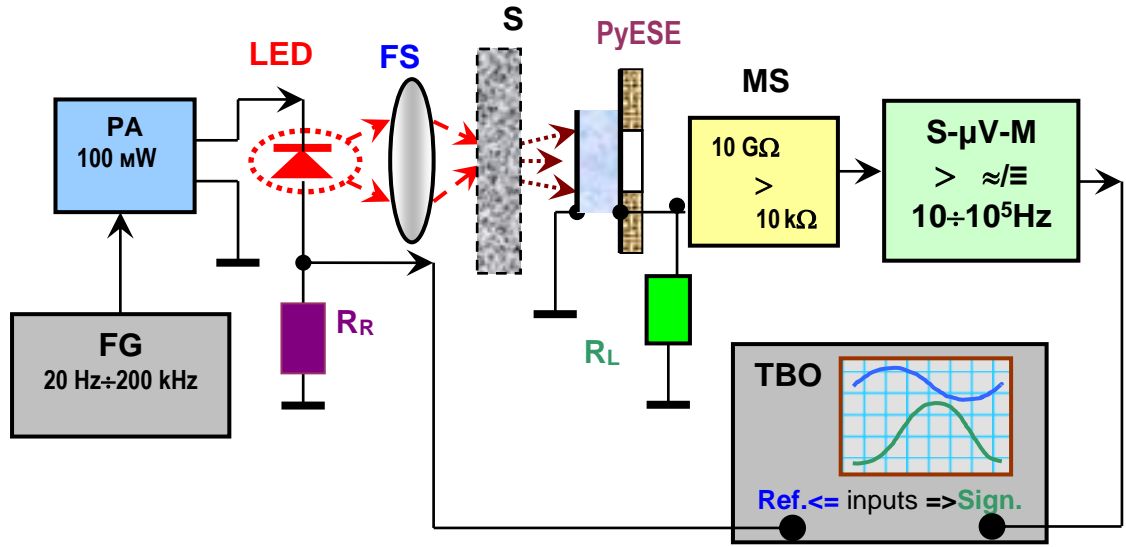

(a)

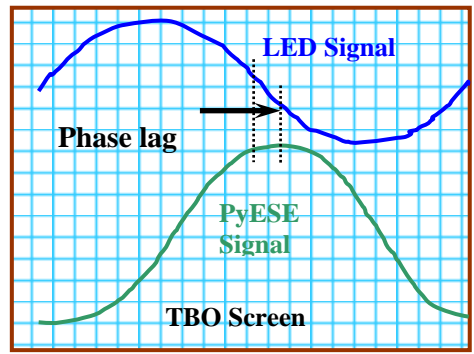

(b)

**Figure 3.** (**a**) - Block diagram of the thermo-wave modulation method.

FG – is the Functional generator (sine, triangle, meander, pulse output voltage form); PA – is the power amplifier (impedance transformer); LED – is the light emitting diode (~ 10 mW, ≈1 μm); FS – is the focusing system; S – is the sample under test; PyESE – is the pyroelectric sensitive element (LiTaO$_3$); R$_L$ – is the load resistor of PyESE; MS – is the matching stage (impedance transformer); S-µV-M – is the selective microvolt-meter 20 Hz – 200 kHz; $R_R$ – is the reference resistor; TBO – is the two-beam oscilloscope.

(**b**) - illustration of phase lag between "primary" heat fluxe (from LED) and "secondary" heat fluxe registered by PyES.

In both HPM and TWM applied, light emitting diode (LED) as radiation heater and the pyroelectric detector as heat sensor (PyES) were used.

The time delay between responses of PyES on "primary" heat fluxe (with no of sample) and "secondary" heat fluxe (with a sample) for HPM is illustrated in the **Fig. 2b.**

The phase lag between voltage on the reference resistor, which is directly proportional to "primary" heat fluxe from LED, and "secondary" heat fluxe registered by PyESE for TWM is illustrated in the **Fig. 3b**.

### 2.3.1. Photo-thermal heat pulse method: Measurements and results

Usually, in photo-thermal HPM the heat pulse duration τ is many times less than the transit time $t_x$ of the heat pulse through the specimen [90 - 92]. The time $t_x$ is usually expressed in the form $t_x = q_x d^2/a_T$, where $d$ is the actual heat travel distance and the value of multiplier $q_x$ depends on the choice of real or extrapolated position of $t_x$ on temperature - time curve for rear interface of the sample [92 - 94]. For $t_x = t_{0.5}$, the time corresponding to 50 % of the maximal temperature increase, $q_x = 1.37/\pi^2 \approx 0.14$, and for $t_x = t_0$, the extrapolated intercept on the time axis at zero temperature rise, $q_x = 0.48/\pi^2 \approx 0.05$ [92]. The choice of time $t_x$ corresponding to a percentage of the maximal temperature increase from 10 to 90% leads to an increase of $q_x$ value from ≈ 0.07 to ≈ 0.3, respectively [94]. Thus, the values of $t_x$ and $q_x$ in the expression for evaluating the diffusivity value $a_T = q_x d^2/t_x$ can be different depending on the choice of "actual point" on sensor response - time curve. If the pulse time τ is comparable in duration with $t_x$, the temperature rise on the rear face of the sample will be retarded (the so called "finite pulse-time effect"), and $q_x$ will be several times higher than the values indicated above [90 - 92].

In the modified HPM proposed and developed by Lang [75] and improved by Yeack et al. [76] to estimate the thermal diffusivity $a_T$, the time $t_{peak}$ between the laser pulse and the peak of the impulse response of the PyES was used: $a_T = q_x d^2/t_{peak}$.

Inasmuch as current response of PyES is proportional to the velocity $dT/dt$ of temperature variation [95 - 97], the maximum of the response-time curve of PyES operating in the current mode corresponds to the maximal slope $dT/dt$ of temperature-time $T(t)$ curve. As a matter of fact, the shape of the pulse response of pyroelectric sensor is determined not only by the thermal properties of both the sample and the sensor [76], but also by the parameters of the electric circuit of the PyES [75, 97], i.e., by the finite rise time of the sample - sensor system on the whole. Therefore, for the case of a one-dimensional heat flow and the use of a PyES, the value $q_x \approx 1/6 \approx 0.17$ was proposed in the Ref. [75], and the value $q_x \approx 1.48$ was established in the Ref. [76].

In the HPM applied **(Fig. 2a)**, the front surface of TEG-CNT-cs sample was exposed to "primary" pulse radiation flux of 100 ms of duration from DC-impulse driven LED. The delayed "secondary" IR-radiation pulse flux from the rear surface of TEG-CNT-cs sample was transformed in a voltage pulse by PyES with IR-window and after amplification, analog-to-digital conversion and PC processing was displayed on the PC monitor screen **(Fig. 2b)**.

The values of thermal diffusivity $a_T$ were estimated from the difference $\Delta t_d$ in time of the maxima $t_m$ of the pyroelectric response to the incident "primary" radiation (with no of sample) and to the delayed "secondary" radiation **(Fig. 2b)**, and known thickness $d$ of the sample (front to rear face distance) at $q_x = 1$: $a_T = d^2/\Delta t_d$.

Measured time $t_m$ of PyES response maximum, the time of the delay $\Delta t_d$ of temperature front diffusion for TEG-CNT-cs samples of different thickness and preparation ways (ECAO or CO), and calculated $a_T$, $b_T$ and $k_T$ effective values are presented in the **Table 2**.

**Table 2.** Samples thicknesses $d$, the time $t_m$ of pyroelectric sensor response maximum, delay time $\Delta t$, and calculated $a_T$, $b_T$ and $k_T$ effective values for different TEG-CNT samples.

| Parameter<br>Sample number & preparation way | $d$, µm | $t_m$, ms | $\Delta t_d$, ms | Diffusi-vity, $a_T$, $10^{-6}$ m²/s | Effusivity, $b_T = C_\rho a_T^{1/2}$, $10^3$ Ws$^{1/2}$/m²K | Thermal conduc-tivity, $k_T = a_T C_\rho$, W/m·K | Correc-ted for G-frct., $k_T^{corr}$, W/m·K | Correc-ted for EG-frct., $k_T^{corr}$, W/m·K |
|---|---|---|---|---|---|---|---|---|
| air (no of sample) | | 392 | 0 | 22 | 5.6·10⁻³ | 0.026 | | |
| 1-E_Chem- CNT | 250 | 427 | 35 | 1.8 | 1.0 | 1.3 | 7.2 | 3.6 |
| 2-Chem-CNT | 300 | 403 | 11 | 8.2 | 2.1 | 6.1 | 50.8 | 25.1 |
| 3-Chem-CNT | 450 | 421 | 29 | 7.0 | 2.0 | 5.2 | 45.2 | 22.6 |
| 4-Chem-Sem | 480 | 421 | 29 | 8.0 | 2.1 | 5.9 | 49.2 | 24.6 |
| 5-Chem-China | 480 | 421 | 29 | 8.0 | 2.1 | 5.9 | 51.3 | 25.7 |
| 6-Elechem-Mikle | 300 | 415 | 23 | 3.9 | 1.5 | 2.9 | 16.1 | 8 |
| Note | | | | | 1) $C_\rho$ of TEG | 1) $C_\rho$ of TEG | 2) $k_T^G$/ /G-frct. | 3) $k_T^{EG}$/ /EG-frct. |

Notes:

1) For $b_T$ and $k_T$ estimation the data of Ref [74] on $C_\rho$ (averaged ≈ 0.74 MJ/m³·K) and $\rho$ (averaged ≈ 1 g/cm³) values of TEG were used.

2) $k_T^{corr}$ are the $k_T$ values corrected for G-fraction (see Table 1). (See in the Discussion)

3) $k_T^{corr}$ are the $k_T$ values corrected for TEG-fraction (see Table 1). (See in the Discussion)

### 2.3.2. Photo-thermo-wave modulation method: Measurements and results

In the Ångström [84, 85] TWM method modified for rod shaped samples [86, 99], the relation of the phase shift $\Delta\varphi$ between emitted and detected temperature waves and diffusivity $a_T$ is given by the expression $\Delta\varphi = d/\lambda_T = d/(2a_T/\omega)^{1/2}$, where $d$ is the inter-probe distance [86] or distance between heating and probing points [99], $\lambda_T = (2a_T/\omega)^{1/2}$ is the thermal diffusion length, $\omega = 2\pi f$ is the angular frequency, $f$ is the cyclic frequency. For $a_T$ the above equality for $\Delta\varphi$ gives the expression $a_T = (\omega/2)(d/\Delta\varphi)^2 = \pi f(d/\Delta\varphi)^2$.

In the modifications of Ångström [84, 85] TWM method for disk shaped samples [87, 88], at given sample thickness (heater-to-sensor distance) $d$ and period $T = 2\pi/\omega$ of temperature wave, a similar expression for $a_T$ was used: $a_T = \pi d^2/T(\Delta\varphi^2)$.

In the TWM applied, the values of thermal diffusivity $a_T$ were calculated from the phase shift $\Delta\varphi$ of the pyroelectric response to the "secondary" radiation relative to the incident "primary" irradiation with a known sample thickness $d$. At that, the above expressions for estimation the value of $a_T$ can be written as $a_T = \pi f(d/\Delta\varphi)^2$.

Under direct oscilloscopic observation, the ratio $\Delta\varphi/2\pi = \Delta t/T$ is valid, and the above expression for $a_T$ can be rewritten as $a_T = (T/4\pi)(d/\Delta t)^2 = (1/4\pi f)(d/\Delta t)^2$, where $T$ and $\Delta t$ can be read-out directly from the oscilloscope screen scale.

In the TWM (**Fig. 3a**), the front surface of TEG-CNT-cs sample is irradiated by sinusoidally modulated with frequency 20 Hz "primary" IR-radiation flux from IR LED driven by functional generator. The sinusoidal "secondary" IR-radiation flux from the rear surface of the sample, which is lagging in phase, is detected by pyroelectric $LiTaO_3$ sensitive element connected with impedance transformer, and after amplification by selective microvolt-meter is displayed on the screen of double-beam oscilloscope together with the current of LED **linearly** related to the intensity of its radiation (**Fig. 3b**).

Measured phase lag $\Delta\varphi$ of pyroelectric sensor response and corresponding time of temperature wave delay $\Delta t$ for TEG-CNT-cs samples of different thickness and preparation way, as well as calculated $a_T$, $b_T$ and $k_T$ effective values are presented in the **Table 3**.

**Table 3.** Samples thicknesses $d$, the phase lag $\Delta\varphi$ of pyroelectric sensor response, delay time $\Delta t$, and calculated $a_T$, $b_T$ and $k_T$ effective values for different TEG-CNT samples.

| Parameter<br>Sample number & preparation way | $d$, µm | $\Delta\varphi$, $2\pi$ | $\Delta t$, ms | Diffusivity, $a_T$, $10^{-5}$ m²/s | Effusivity, $b_T=C_\rho a_T^{1/2}$, $10^3$ Ws$^{1/2}$/m²K | Thermal conductivity, $k_T=a_T C_\rho$, W/m·K | Corrected for G-fract., $k_T^{corr}$, W/m·K | Corrected for EG-fract $k_T^{corr}$, W/m·K |
|---|---|---|---|---|---|---|---|---|
| 0-air (no of smp.) | | 0 | 0 | 2.2 | 5.6·10⁻³ | 0.026 | | |
| 1-E-Chem-1CNT | 250 | 0.08 | 4 | 1.6 | 3.0 | 11.8 | 65.6 | 32.8 |
| 2-Chem-CNT | 300 | 0.05 | 2.5 | 5.5 | 5.5 | 40.8 | 340 | 170 |
| 3-Chem-CNT | 450 | 0.06 | 3 | 8.8 | 6.9 | 65.1 | 566 | 283 |
| 4-Chem-Sem | 500 | 0.055 | 2.75 | 13.5 | 8.5 | 100 | 832.5 | 416.3 |
| 5-Chem-China | 450 | 0.05 | 2.5 | 13 | 8.4 | 96.2 | 836.5 | 418.3 |
| 6-Elechem-Mikle | 300 | 0.08 | 4 | 2.2 | 3.5 | 16.3 | 90.6 | 45.3 |
| Note | | | | 1) $C_\rho$ of EG | 1) $C_\rho$ of EG | 2) $k_T$/G-fract. | 3) $k_T^{TEG}$/EG-fract. | |

Notes:

1) For $b_T$ and $k_T$ estimations, the data of Ref [74] on values of $C_\rho \approx 0.74$ MJ/m$^3$·K (averaged) and $\rho \approx 1$ g/cm$^3$ (averaged) of EG were used.

2) $k_T^{corr}$ are the $k_T$ values corrected for G-fraction (see Table 1). (See in the Discussion)

3) $k_T^{corr}$ are the $k_T$ values corrected for TEG-fraction (see Table 1). (See in the Discussion)

## 3. DISCUSSION

### 3.1. Thermal diffusivity, effusivity and conductivity: Some comparisons

The diffusivity $a_T$ values obtained for different TEG-CNT-cs samples by time delay T-pulse and phase lag T-wave measurements (**Tables 2** and **3**, respectively) were found in the range $(10^{-6} - 10^{-4})$ m$^2$/s.

The effusivity $b_T$ values for different TEG-CNT-cs samples, estimated using the data on the density and heat capacity of thermally expanded (exfoliated) graphite collected in the Ref. [74], were found in the range $(10^3 - 10^4)$ Ws$^{1/2}$/m$^2$K.

These $a_T$ and $b_T$ values for TEG-CNT composites are in the ranges limited by $a_T$ and $b_T$ values known for the quartz, SiO$_2$, ($a_T \sim 10^{-6}$·m$^2$/s and $b_T \approx 10^3$ Ws$^{1/2}$/m$^2$K), and for Si, GaAs, Al, Cu ($a_T \sim 10^{-4}$·m$^2$/s and $b_T \sim 10^4$ Ws$^{1/2}$/m$^2$K).

Known for today $a_T$ values for certain C-materials are:

- $(0.7 - 10) \cdot 10^{-5}$ m$^2$s$^{-1}$ for MWCNT of different length (12-46 μm) [60];
- $6.2 \cdot 10^{-6}$ m$^2$s$^{-1}$ for MWCNT yarns [62];
- $4.5 \cdot 10^{-5}$ m$^2$s$^{-1}$ and $8.2 \cdot 10^{-6}$ m$^2$s$^{-1}$ for highly aligned MWCNT sheets for along and perpendicular to the CNT alignment, respectively [62, 63];
- $\approx 1 \cdot 10^{-4}$ m$^2$s$^{-1}$ for "nonwoven" CNT-sheets made from dry spinning CNTs [16];
- 6.5, 1.9 and $1.3 \cdot 10^{-4}$ m$^2$s$^{-1}$ for 4-, 8- and 16- layered graphene, respectively [64];
- from $3.5 \cdot 10^{-5}$ to $0.5 \cdot 10^{-5}$ m$^2$/s for graphene nanoplatelets, under the density increase from 0.1 to 1 g/cm$^3$ [65].

Thus, the $a_T$ values of different TEG-CNT-cs samples (Tables 2, 3) are comparable with $a_T$ values known for multiwall CNT sheets [62, 63], as well as with $a_T$ values known for graphene multilayers [64] and nanoplatelets [65] (see the numbers above).

The range of $b_T$ values for TEG-CNT-cs covers the value of 6 10$^3$ Ws$^{1/2}$/m$^2$K, known for the polyvinylidene fluoride (PVDF)-graphene nanocomposite with 40 wt.% of graphene nanoplatelets [100], and is close to the value of $\approx$ 2 10$^4$ Ws$^{1/2}$/m$^2$K, known for the graphene nanoplatelet filler in the polyvinyl chloride (PVC)/graphene nanocomposite [101].

Corresponding thermal conductivity $k_T$ values of TEG-CNT-cs were estimated using data [74] on the density and heat capacity of thermally expanded (exfoliated) graphite. Obtained $k_T = a_T C_\rho$ values for different TEG-CNT-cs samples vary in the range (1 - 100) W/m·K (**Tables 2, 3**).

Known $k_T$ values for certain regular graphene derivative materials are:

- close to 5000 W/m K for monolayer graphene "flakes" (see [20] and Refs [5] and [110] therein);
- in the range (1770 - 2507) W/m K depending on the porosity degree (20 – 35) % for graphene nanosheets at 323 K [48];
- $\approx$ 2500 W/m K for graphene grown on the SiN-membrane substrate (see [20] and Ref. [113] therein) and $\approx$ 600 W/m K for micromechanically exfoliated graphene deposited on SiO$_2$-substrate (see [20] and Refs [113] and [114] therein);
- $\approx$ 600 W/m K for graphene supported by SiO$_2$, $\approx$ 160 W/m K for SiO$_2$-encased graphene, and $\approx$ 80 W/m K for supported 20-nm-wide graphene nanoribbons (see [102] and Refs [33], [40] and [34] therein);
- from 40 W/m K to 90 W/m K for graphene laminate films on polyethylene terephtalate substrate depending on the average size and the alignment of graphene flakes and degree of compression [31].

For certain irregular graphene derivative materials the following $k_T$ values were reported:
- 3.87 ± 0.28 W/mK for graphene/epoxy resin 10 vol % composites [47];
- 2.06 and 10 W/m K for graphene/PVDF composite at a filler content of 20 and 25 wt% respectively (Ref. [52] in [30]);
- 5.8 W/m K for 20 wt% silane-functionalized graphene/epoxy resin composite (Ref. [71] in [30]);
- 6.44 W/mK for 25 wt% of graphene nanoparticles/epoxy resin composite (Ref. [72] in [30]);
- 33.54 W/m K for graphene aerogel/epoxy resin composite (Ref. [73] in [30]).

For C-C-composites, the considerable difference of the in-plane $k_T^i$ and through-plane $k_T^t$ values is characteristic. Thus, for the hot-pressed 15 wt% hybrid graphene nanoparticles/cellulose composite $k_T^i$ = 41 W/m K and $k_T^t$ = 1.2 W/m K ($\approx$ 34 times ratio) [30], for commercial flexible graphite $k_T^i$ = 43 W/m K and $k_T^t$ = 3 W/m K ($\approx$ 14 times ratio) and for graphite nanoplatelets paper $k_T^i$ = 180 W/m K and $k_T^t$ = 1.3 W/m K ($\approx$ 138 times ratio) (see [26] and Refs [53] and [71] therein).

The comparison with above data shows that the maximal $k_T$ values of TEG-CNT-cs samples ~100 W/m K (**Tables 2**, **3**) are comparable with $k_T$ values known for encased graphene and supported graphene nanoribbons (see [102]), as well as for graphene laminate films (see [31]) (the numbers are presented above).

The range of $k_T$ values for different TEG-CNT-cs samples (**Tables 2**, **3**) overlaps with the ranges of $k_T$ values known for graphene-based polymer composites (see [30]) and for in-plane and through-plane thermal conductivity of C-C-composites (see [26] and the numbers presented above).

The relatively low $k_T$ values of TEG-CNT-cs are apparently associated with an increased interlayer spacing between the individual graphene sheets and with a weak thermal coupling between the irregularly stacked graphene species in the EG flakes of TEG-CNT-cs (see **Fig. 1**).

The values of $a_T$, $b_T$ and $k_T$ can be compared with those estimated using an expression similar to that proposed for predicting the thermal conductivity of two-phase polymer composites [103] and applied to estimate the diffusivity and effusivity of graphene-polymer nanocomposites [101]:

$$h_{TE}^{\zeta} = v\, h_{TG}^{\zeta} + (1 - v)\, h_{TA}^{\zeta}, \qquad (1)$$

where $h_{TE}$ are $a_{TE}$, $b_{TE}$ or $k_{TE}$ estimated effective values, $v$ and $(1 - v)$ are the <u>volume</u> fractions of graphene (G) with $h_G$ and air (A) with $h_A$, respectively, and $-1 < \zeta < 1$ is the fitting parameter that depends on the model choise ($\zeta = -1$ and $\zeta = 1$ are the limits for series ("vertical" [103]) and <u>parallel</u> model, respectively [103]).

The values of $a_{TE}$ estimated from (1) at $v = 0.25$ for the parallel ($\zeta = 1$, $a_{TE} \approx 4.2 \cdot 10^{-5}$ m$^2$/s), mixed ($\zeta = 0.5$, $a_{TE} \approx 3.6 \cdot 10^{-5}$ m$^2$/s) and series ($\zeta = -1$, $a_{TE} \approx 2.7 \cdot 10^{-5}$ m$^2$/s) models, differ insignificantly in view of the not very large difference between the values of $a_T$ of graphene nanoplatelets (from $0.5 \cdot 10^{-5}$ m$^2$/s to $3.5 \cdot 10^{-5}$ m$^2$/s) [65] and air ($2.2 \cdot 10^{-5}$ m$^2$/s, **Tables 2**, **3**) and are within the obtained $a_T$ values (see the **Tables 2**, **3** and the numbers above).

From (1), at $v = 0.25$ for the parallel model ($\zeta = 1$) the value $b_{TE} \approx 2.5 \cdot 10^3$ Ws$^{1/2}$/m$^2$K is within the interval of obtained $b_T$ values (see **Tables 2, 3**), for the mixed model ($\zeta = 0.5$) $b_{TE} \approx 6 \cdot 10^2$ Ws$^{1/2}$/m$^2$K $< b_T$, and for the series model ($\zeta = -1$) $b_{TE} \approx 0.8 \cdot 10^1$ Ws$^{1/2}$/m$^2$K $<< b_T$. These differences in $b_{TE}$ are associated with a significantly lower $b_T$ value of air (5.6 Ws$^{1/2}$/m$^2$K, **Tables 2**, **3**) than the $b_T$ value of graphene nanoflakes ($\approx 2 \cdot 10^4$ Ws$^{1/2}$/m$^2$K) [101].

From (1), at $v = 0.25$ for the parallel model ($\zeta = 1$) the value $k_{TE} \approx 25$ W/m·K is within the range of obtained $k_T$ values (see **Tables 2**, **3**), for the mixed model ($\zeta = 0.5$) $k_{TE} \approx 6$ W/m·K $< k_T$, and for the series model ($\zeta = -1$) $k_{TE} \approx 0.03$ W/m·K $<< k_T$. These differences in $k_{TE}$ are due to the fact that the air $k_T$ value (0.024 W/m·K, **Tables 2**, **3**) is several orders of magnitude lower than the $k_T$ value of both few-layer graphene [104, 105] and graphene in thermal contact with SiO$_2$ (see [20] and Ref. [114] therein, [31], [102] and Refs [33], [40] and [34] therein, as well as [104, 106, 107]).

Such significant discrepancies, as noted earlier [103], are typical for most polymer composites, when the experimental values of $k_T$ can be less than the calculated value of $k_{TE}$ for the parallel model and greater than the calculated value of $k_{TE}$ for the series model.

The different inequalities for $a_{TE}$ and $b_{TE}$ obtained for the different models seem to correspond to the different physical meaning of $\alpha_T = k_T/C_\rho$ and $b_T = (k_T C_\rho)^{1/2} = a_T^{1/2} C_\rho$, namely the ability of the material to spread heat and to exchange heat with its surroundings, respectively.

Estimation of thermal conductivity of dense-packed TEG- and G- components in the low density TEG-CNT-cs was performed by the correction [11, 57] for low TEG- and G-fractions (see **Table 1**). These corrected $k_T^{corr}$ values can be estimated by applying the parallel thermal conductance model used in Ref. [60] for MWCNT films in the form

$$k_{TE} = v k_T^{corr} + (1 - v) k_{TA}, \qquad (2)$$

where $k_{TE}$ are the effective values of $k_T$, $v$ and $(1 - v)$ are the volume fractions of dense-packed TEG component with $k_T^{corr}$ and air (A) with $k_{TA}$, respectively.

From (2), at $v$ values given in the **Table 1**, the $k_T^{corr}$ values for dense-packed TEG- and G-components in the TEG-CNT-cs were obtained in the range (3 - 500) W/m.K and (6 - 1000) W/m.K, respectively (see **Tables 2, 3**).

The minimal $k_T^{corr}$ values characteristic of TEG-CNT-cs samples obtained by ECAO are comparable with $k_T$ values known for graphene/thermoplastic polymer composites [30] and with $k_T^t$ values known for C-C-composites [26] (see the numbers above).

The maximal $k_T^{corr}$ values characteristic of TEG-CNT-cs samples obtained by CO are comparable with $k_T$ values known for exfoliated graphene deposited on $SiO_2$-substrate [20] and for graphene supported by $SiO_2$ [102] (see the numbers above).

The difference in the values of $a_T$, $b_T$ and $k_T$ for the TEG-CNT-cs samples obtained by ECAO and by CO may be associated with the difference of electrochemical and chemical processes during the preparation of TEG-CNT-cs, which resulting in different density even after the same rolling, affects the state of G-fraction and, consequently, affects the thermophysical parameters of ECAO and CO samples.

### 3.2. Thermophysical parameters with density correlations

**Figure 4** show the estimated diffusivity $a_T$ and effusivity $b_T$ values in comparison with the density values obtained by HPM **(Fig. 4a)** and TWM **(Fig. 4b)** for TEG-CNT-cs obtained by ECAO and by CO.

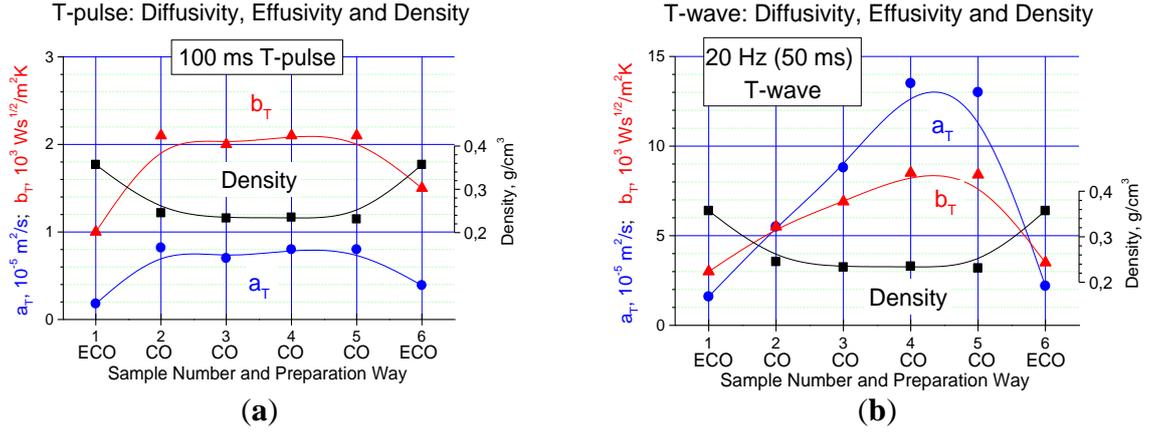

**Figure 4.** The relationship between the values of density, diffusivity $a_T$ and effusivity $b_T$ obtained by HPM (**a**) and TWM (**b**) for TEG-CNT composites obtained by ECAO (samples 1 and 6) and by CO (samples 2 - 5);

Several points should be noticed:

(1) One can see that ECAO samples with 1 % of CNTs have the maximal density values and CO samples with 2 % of CNTs have the density values in one and a half time lower (**Fig. 4a,b**).

(2) One can notice that CO samples with minimal density have maximal $a_T$ and $b_T$ values and ECAO samples with maximal density have minimal $a_T$ and $b_T$ values (**Fig. 4a,b**).

(3) For both $a_T$ and $b_T$ data obtained by HPM (**Fig. 4a**) and by TWM (**Fig. 4b**) the correlations value - density have the similar character.

Therefore, the evidence is presented for correlation of diffusivity $a_T$ and effusivity $b_T$ values with the mass density of the TEG-CNT-cs obtained by ECAO and CO methods.

The trend of thermal diffusivity decrease at density increase observed for TEG-CNT-cs (**Fig. 4a,b**) is known for graphene nanoplateles [65]: $a_T$ decreases from 3.5 to 0.5 $10^{-5}$ m$^2$/s (7 times of ratio) at density increases from 0.1 to 1 g/cm$^3$ (10 times of ratio). In this case, the relatively small increase in $k_T$ from 2.5 to 3.5 W/mK ($\approx$ 1.5 times) shows that the considerable decrease in $a_T = k_T/C_\rho$ is associated with higher $C_\rho = \rho c_p$ values in denser samples.

In the case of TEG-CNT-cs (**Fig. 4a,b**), there is a significant change in the values of $a_T$ (from 4 to 7 times) at a relatively small change in density ($\approx$ 1.5 times), but the trend is the same as for graphene nanoplatelets [65]: decrease in $a_T$ with increasing density.

It should also be noted that such trend for the TEG-CNT-cs and for the graphene nanoplates [65] is inverse to the established one for the graphene nanosheets [48], where at 323 K the porosity decrease from 35 to 20% leads to the increase of $k_T$ from 1770 to 2507 W/m·K, as well as for the graphene laminate [31], where under compression the density change from 1.0 to 1.9 g/cm$^3$ leads to almost two-fold increase of the $k_T$ value.

At the same time, for few-layer graphene, the thermal conductivity decreases from ≈ 2800 to ≈ 1300 W/m·K as the number of layers increases from 2 to 4 [104], which is associated with the appearance of additional phonon scattering channels due to the interlayer cross-plane coupling arising owing to the transition from 2-D to 3-D structure [104].

For SiO$_2$ supported single-layer graphene, the interaction with the substrate suppresses the thermal conductivity to 600 W/m·K owing to the phonon leakage through the graphene-substrate interface [106].

However, for multilayer graphene supported by amorphous SiO$_2$, the thermal conductivity increases with an increase in the number of layers, which is caused by a decrease in support-related phonon scattering owing to a decrease in interfacial phonon leakage [107].

As distinct from spatially ordered C-materials, in the heterogeneously structured ones, such as graphene nanosheets-CNT composites [48], graphene laminate [31], defective graphene [108] and graphene nanoribbons [102], the flake boundaries rather than intrinsic properties of the graphene and few-layer graphene were considered as a limiting factor for heat conduction [31, 102, 108].

Therefore, the considerable difference in the values of $a_T$ (from 4 to 7 times) and $b_T$ (from 2 to 3 times) for ECAO and CO samples of TEG-CNT composites (**Fig. 4a,b**) can be associated with higher quality of thermal contacts between few-layer graphene platelets in EG flakes of less dense CO samples in comparison with that of the denser ECAO samples.

It should be mentioned that similar decrease of $k_T$ values ($k_T = a_T C_\rho$) from ≈ 1800 to ≈ 400 W/m K (in 4.5 times) was found for suspended graphene when the density of defects was changed from ≈ 2·10$^{10}$ to ≈ 1.8·10$^{11}$ cm$^{-2}$ (in 9 times) by electron beam irradiation [108].

Thus, most likely, the difference in the state of defects, in particular their amount and geometry, as well as edge and interface roughness, in TEG-CNT-cs, denser obtained by ECAO and less dense obtained by CO, is associated with the electrical current treatment at ECAO and its absence at CO.

**3.3. Mean free path and relaxation time with graphite fraction correlations**

The mentioned above expressions for thermal diffusivity $a_T = k_T/C_\rho$ and effusivity $b_T = (k_T C_\rho)^{1/2} = C_\rho a_T^{1/2}$ include the thermal conductivity $k_T$ and the volume heat capacity $C_\rho = c_p \rho$, where $\rho$ is the density. Indeed, the higher the density, the higher the $C_\rho$ value. However, an increase in $a_T$ by several times cannot be solely due to a decrease in the heat capacity only with a decrease in density by one and a half times (see **Fig. 4**).

Taking into account that in the non-metallic solids [56] $k_T = C_\rho v_p l_p/3 = C_\rho v_p^2 \tau_p/3$, where $v_p$, $l_p$ and $\tau_p$ are the averaged phonon group velocity (≈ speed of sound), mean free path (MFP) and relaxation time (RT), respectively, one can obtain $a_T = v_p l_p/3 = v_p^2 \tau_p/3$ and $b_T = C_\rho(v_p l_p/3)^{1/2} = C_\rho v_p (\tau_p/3)^{1/2}$. Therefore, in the spatially ordered materials, including C–C- ones [9, 11, 31, 63], namely the MFP or RT (along with $C_\rho$ for $b_T$) are the factors limiting the $a_T$ and $b_T$ values.

It can be assumed that in the case of TEG-CNT-cs the changes of $a_T$ and $b_T$ are also associated with changes of the average values of MFP or RT.

For TEG-CNT-cs, using the estimated range of $a_T$ values $(10^{-6} - 10^{-4})$ m$^2$/s (see **Fig. 4**) and $v_p$ values known for single-crystalline graphite $v_p \approx (1.5 - 22) \cdot 10^3$ m/s [109] and for strained graphene $v_p \approx 10^4$ m/s in average for different directions [98], one can estimate the effective values of MFP $l_p = 3a_T/v_p$ and RT $\tau_p = 3a_T/v_p^2$. The relationship between the values of mean free path $l_p$, relaxation time $\tau_p$ and the graphite (G-) fraction for samples of TEG-CNT-cs obtained by ECAO and CO is illustrated in the **Figure 5**.

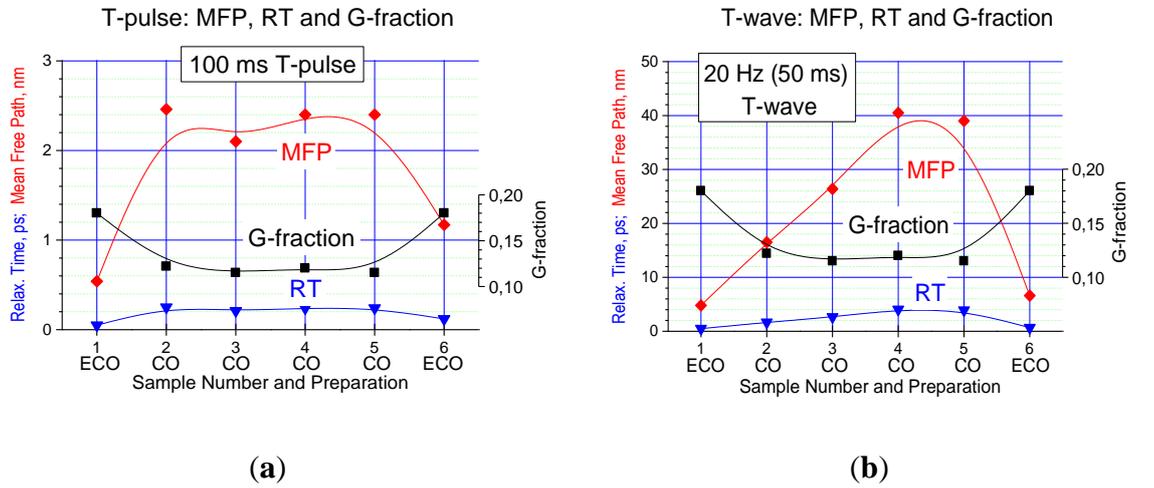

(**a**) (**b**)

**Figure 5.** The relationship between the values of graphite (G-) fraction, mean free path (MFP), and relaxation time (RT) obtained by HPM (**a**) and TWM (**b**) for ECAO and CO samples of TEG-CNT composites;

The ranges of both $l_p$ and $\tau_p$ values for TEG-CNT-cs, $l_p = (0.5 - 40)$ nm and $\tau_p = (0.05 - 4)$ ps (**Fig. 5a,b**) are close to those estimated for phonon MFPs and RTs in defected (~ 1 %) graphene [110], (1 - 100) nm and (0.2 - 10) ps, respectively.

To refine the relationship between RT and MFP, following [111], one can use the Ziman formula [112] for the phonon scattering rate

$$1/\tau_B = (v_p/D)((1 - p)/(1 + p)), \qquad (3)$$

where $\tau_B$ is the phonon lifetime, $D$ is the characteristic size of nanostructure, and p is the specularity parameter defined as the probability of specular scattering at the boundary [111].

Using (3), the estimation of $D = v_p \tau_B((1 − p)/(1 + p))$ for TEG-CNT composites at $\tau_B \approx 1$ ps (see **Fig. 5b**), $v_p = 10^4$ m/s and p = 0.3 (rough interfaces) gives $D \approx 5$ nm, which is within the range of $l_p$ values obtained from $a_T$ measurements (see the numbers above).

The obtained $l_p$ values are much smaller than MFP known for suspended graphene ($\approx 600$ nm) and are smaller of that known for graphene supported on $SiO_2$ ($\approx 100$ nm) [102].

Maximal $l_p$ value $\approx 40$ nm is comparable with MFP known for MWCNTs ($\approx 20$ nm) [60] and for graphene nanoribbons ($\approx 20$ nm of width) supported on $SiO_2$ ($\approx 20$ nm) [102]. This $l_p$ value and corresponding maximal RT value $\tau_p \approx 4$ ps are in the ranges of MFPs, (10 - 100) nm, and RTs, (1 - 10) ps, estimated for low-frequency acoustic ZA, TA and LA phonons (below 3, 5 and 10 THz, respectively) in double- and mono-vacancy-defected ($\sim 1$ %) graphene [110].

Minimal $l_p$ value $\approx 0.5$ nm is much smaller than MFP known for MWCNTs with defects $\approx 4$ nm [9, 113] and is close to MFP estimated for high-frequency acoustic ZA and TA phonons (at 13 and 23 THz, respectively) in mono-vacancy-defected (> 1 %) graphene ($\approx 0.4$ nm) [110]. Corresponding minimal $\tau_p$ value $\approx 0.05$ ps is much smaller than the minimal RT values ($\approx 0.2$ ps) estimated for high-frequency acoustic TA (near 20 THz) and LA (near 20 and 40 THz) phonons in mono-vacancy-defected ($\sim 1$ %) graphene [110].

Thus, reduced $l_p$ and $\tau_p$ values in the TEG-CNT-cs can be associated with strong phonon scattering on the point and extended defects and on the EG-flake rough edges as well as on the graphene-graphene interfaces.

In this regard, it should be noted that, according to Cai et all. [114], the thermal conductivity of supported CVD graphene (370 + 650/-320) W/m K near room temperature are significantly lower than those of suspended CVD graphene (2500 + 1100/-1050) W/m K near 350 K. Almost simultaneously it was shown by Jang et all. [105] that for graphene and ultrathin graphite (thickness from 1 to $\approx$20 layers) encased within silicon dioxide $SiO_2$ the characteristic penetration distance of surface-induced disturbances into the adjacent graphene layers is approximately 2.5 nm (7 layers) at room temperature. Later on, the molecular dynamics simulations, combined with the spectral energy density analysis performed by Qiu and Ruan [115], predicted a significant decrease in RT for acoustic ZA-, TA-, and LA-phonons from (5 - 45) ps up to (1 - 10) ps, as well as for optical ZO, TO and LO phonons from (2 - 30) ps to (0.5 - 10) ps when the state of graphene changes from suspended to supported on amorphous $SiO_2$ and a strong bond appears between graphene and the surface of the $SiO_2$ substrate.

In this respect, the state of the neighbouring graphene layers located between the others in G-flakes of TEG-CNT-cs can be considered as similar to that of the graphene supported on $SiO_2$ [102, 107, 115], or similar to that of the few-layer graphene encased within $SiO_2$ [105], with a corresponding reduction of MFP and decrease of RT.

It is seen from **Fig. 5a,b**, that TEG-CNT-cs samples with high G-fraction obtained by ECAO have the lower $l_p$ and $\tau_p$ values, and TEG-CNT-cs samples with low G-fraction obtained by CO have the higher $l_p$ and $\tau_p$ values.

The aforementioned decrease in thermal conductivity for few-layer graphene from ≈ 2800 to ≈ 1300 W/m K with an increase in the number of layers from 2 to 4, associated with an increase in the scattering of low-frequency phonons at interlayer boundaries [104], may also reflect a corresponding decrease in MFP and RT values.

Thus, low and high values of MFP and RT in samples of TEG-CNT-cs with high and low G-fraction obtained by ECAO and CO, respectively, can be associated with a large number of graphene layers in G-flakes of samples electrically treated with ECAO and with a small amount of graphene layers in G-flakes of electrically untreated CO samples.

It should be noted here that the calculations from first principles [116] and based on the exact numerical solution of the Boltzmann equation for phonons [117] showed a significant increase in the lattice thermal conductivity of G-flakes with an increase in their linear size, which is especially sharp in the near-micrometer range. It was further found for graphene laminate [31] that a decrease in the average size of G-flakes by 25% leads to an almost 100% decrease in the thermal conductivity for both uncompressed (1.0 g/cm$^3$) and compressed (1.9 g/cm$^3$) of samples.

Thus, for TEG-CNT-cs, the increase in the G-fraction is most likely associated with an increase in the number of graphene layers, their greater deformation and decrease in the average size of G-flakes. Therefore, the low values of MFP and RT for TEG-CNT-cs are a consequence of an increase in phonon scattering by internal defects, interlayer boundaries, and G-flake edges.

### 3.4. On the difference of diffusivity values by HPM and TWM

For investigated TEG-CNT-cs samples (see Table 1) is characteristic the significant differences of diffusivity $a_T$ values, which were obtained by HPM at 100 ms duration of temperature wave packet [the $a_T$ range ($10^{-6}$ - $10^{-5}$) m$^2$/s], and which were obtained by TWM at 50 ms period (20 Hz) of single sine temperature wave [the $a_T$ range ($10^{-5}$ - $10^{-4}$) m$^2$/s] (compare the scales in **Fig. 4a** and **4b**).

The difference of $a_T$ values in the limits of every range can be related to the stronger and/or weaker thermal coupling (due to interface thermal resistance) between the individual irregularly stacked EG components in different TEG-CNT-cs samples. The degree of this thermal coupling is affected by multitude of point and extended thermal contacts between the EG and CNT species for TEG-CNT-cs samples of different density and so of different porosity degree of samples obtained by ECAO and CO (see **Table 1**).

The difference in the ranges of $a_T$ obtained by the HPM and the TWM may be due to the difference in the manner of sample excitation in these methods [118].

**For HPM**, the diffusivity value could be underestimated due to the finite pulse time effect, which leads to the retardation of a transient temperature rise at the rear sample surface when the pulse time $\tau$ becomes to be comparable to the characteristic heat transit time $t_c = L^2/\pi^2\alpha$ [90 - 92].

It was shown that for a square wave pulse at $0.01 < \tau/t_c < 0.1$ the $t_{1/2}/t_c$ value changes insignificantly and is not far from 1.37, but $\tau/t_c$ increase from 0.1 to 2 leads to appreciable $t_{1/2}/t_c$ increase from 1.4 to $\approx 2.5$ [91]. In this case, the corresponding ratio $\tau/t_c$ could be achieved not only by changing $\tau$, but also by choosing $t_c \sim L^2$.

In the conditions of HPM the process of spreading of temperature wave packet is defined by the effective thermal diffusion length, $\mu_T = 2(a_T t)^{1/2}$, which is similar to the thermal diffusion length $\lambda_T = (2a_T/\omega)^{1/2}$ working in the case of harmonic thermal waves [118]. The dependence of $\lambda_T$ on the frequency $\omega = 2\pi f$ defines not only the depth of thermowave probing [86, 87, 88, 97, 99, 118, 119], but the velocity of temperature waves in the packet.

For phase velocity $V_{ph} = \lambda f$ of temperature wave with wavelength $\lambda = 2\pi\lambda_T$ this gives $V_{ph} = 2(\pi a_T f)^{1/2} = 2(\pi a_T/T)^{1/2}$ ($T$ is the period). For group velocity $V_{gr} = 1/[\partial(1/\lambda_T)/\partial\omega]$ of temperature wave packet with the frequencies in the range $(f, f + \Delta f)$ this gives $V_{gr} = 4(\pi a_T f)^{1/2} = 2V_{ph}$ [118]. Furthermore, $\lambda_T$ value defines the attenuation of temperature wave amplitude by multiplier $\exp(-x/\lambda_T)$, where $x$ is the distance from excitation point [118].

Thus, in the temperature wave packet, the high-frequency temperature waves propagate faster and attenuates stronger than the low-frequency ones and the time of the heat diffusion should be frequency and thickness dependent. In particular, by experimental investigation of finite pulse time effect for industrial pure cupper [120] it was revealed that with decreasing sample thickness from 2 to 0.5 mm (in 4 times) the diffusivity value obtained at constant $\tau$ decreases from $1.2 \times 10^{-4}$ to $0.4 \times 10^{-4}$ m$^2$/s (in 3 times), which is associated with the approach of $t_c$ to $\tau$ due to decrease of $t_c$ value.

**For TWM**, at given $a_T$ value and thickness $d$ of the sample, under $f$ increase, the dependence $\lambda_T(f) = (a_T/\pi f)^{1/2}$ leads to the transition of sample operation as thermally thin one ($d \ll \lambda_T$) at $f \ll a_T/\pi d^2$ to thermally thick one ($d \gg \lambda_T$) at $f \gg a_T/\pi d^2$ (here $d$ is the sample thickness) [97, 118, 119, 121].

The transition frequency $f_t$ and thickness $d_t$ can be estimated from the equality $\lambda_T = (a_T/\pi f_t)^{1/2} \approx d_t$ [97, 121]. Thus, the variation of $d$ at $f$ = const and the variation of $f$ at $d$ = const are equivalent for changing the sample operation conditions. For $d \approx 400$ μm, at $a_T = 10^{-6}$, $10^{-5}$ and $10^{-4}$ m$^2$/s (the case of TEG-CNT-cs, see the **Tables 2**, **3**), on can obtain $f_t$ = 2, 20 and 200 Hz, respectively.

In the Ref. [122] the following expression for effective diffusivity value $a_T^{eff}$ of in series two-layer system with variable thicknesses of the layers has been obtained:

$$a_T^{eff} = 1/[x^2/a_1 + (1 - x)^2/a_2 + x(1 - x)(\kappa/a_1 + 1/\kappa a_2)]. \tag{4}$$

Here the indices 1 and 2 correspond to the materials of the layers 1 and 2, $x = d_1/d$ is the thickness fraction of material 1 in the sample ($d = d_1 + d_2$), and $\kappa = k_1/k_2$.

The expression (4) clearly demonstrates that for two-layer system the $a_T^{eff}$ value depends not only on relative thickness fraction of materials and diffusivities $a_{1,2}$ of the layers 1 and 2, but on the ratio of their thermal conductivities $\kappa$ [122]. Later on, in the Refs. [121, 123] the expression (4) was reconsidered and it was shown that for thermally thick two-layer in series system the effective diffusivity value $a_T^{eff}$ is $\omega$-dependent.

The value of $a_T^{eff}$ for the case when both layers are thermally thick was described by the expression [121]

$$d/\sqrt{a_T^{eff}} = d_1/\sqrt{a_1} + d_2/\sqrt{a_2} + (2/\omega)^{1/2} \ln\{[1 + (a_1/a_2)^{1/2}(k_2/k_1)]/2\}. \tag{5}$$

In the Ref. [123], the value of $a_T^{eff}$ for the case when one layer is in the thin and the other in the thick thermal conditions was described by the expression

$$d/\sqrt{a_T^{eff}} = d_1/\sqrt{a_1} + d_2/\sqrt{a_2} + (2/\omega)^{1/2}\ln\{(a_S^{1/2}/2k_S)(k_1/a_1^{1/2})[1 + (a_1/a_2)^{1/2}(k_2/k_1)]\}, \tag{6}$$

where $\alpha_S$ and $k_S$ are related to the total sample [123].

Taking into account the relationship between effusivity, diffusivity and thermal conductivity, $b_T = k_T/a_T^{1/2}$, the expression (6) was represented [123] as

$$d/\sqrt{a_T^{eff}} = d_1/\sqrt{a_1} + d_2/\sqrt{a_2} + (2/\omega)^{1/2}\ln[(b_1 + b_2)/2b_S], \tag{7}$$

where $b_S$ is related to the total sample.

It follows from (5) and (7) that $\omega$-dependence of the $a_T^{eff}$ take place at $(a_1/a_2)^{1/2}(k_2/k_1) \neq 1$ in the case of (5) and at $(b_1 + b_2)/2b_S \neq 1$ in the case of (7).

At $\omega \rightarrow \infty$, the dependences (5 - 7) transform to the expression $d/\sqrt{a_T^{eff}} = d_1/\sqrt{a_1} + d_2/\sqrt{a_2}$ proposed in the Ref. [124] for combined diffusivity of two-layer system.

The dependences $a_T^{eff}(\omega)$ given by (5 - 7) can be represented in the form

$$a_T^{eff}(\omega) = d^2/[c_0 + c/\omega^{1/2}]^2, \tag{8}$$

where the constants $c_0$ and $c$ depend on the thicknesses $d_{1,2}$, diffusivities $a_{1,2}$ of the layers 1 and 2 and $\kappa$ value.

In the low frequency limit ($\omega \rightarrow 0$) the $a_T^{eff}$ value tends to zero. At intermediate frequencies, the $a_T^{eff}(\omega)$ can be decreasing or increasing function of $\omega$ depending on the $c$ sign ($c < 0$ or $c > 0$), which is defined by the sign of the logarithmic term in (5 - 7), which, in turn, is defined by the values $(a_1/a_2)^{1/2}(k_2/k_1)$ in (5) and $(b_1 + b_2)/2b_S$ in (7) ($< 1$ or $> 1$). In the high frequency limit ($\omega \rightarrow \infty$) $a_T^{eff} = d^2/c_0^2 = d^2/(d_1/\sqrt{a_1} + d_2/\sqrt{a_2})^2$ and $\alpha_T^{eff}$ is minimal at $\kappa \gg 1$ or maximal at $\kappa \ll 1$ and is $\omega$-independent.

The $a_T^{eff}(\omega)$ dependence (5) has been used in the Ref. [121] to explain the change from 3 to 100 times (by one order of the value on average) in the $a_T^{eff}$ value of two-layer systems based on different materials with $\kappa > 1$ and $\kappa \ll 1$ (see the Figs 2 - 5 and the corresponding Refs in [121]).

Subsequently, in the Ref. [125], the existence of finite interface thermal conduction (interface thermal resistance) was taken into account. For $a_T^{eff}$ of a thermally thin two-layer system the following expression was obtained [125]:

$$a_T^{eff} = d^2/[d_1^2/a_1 + d_2^2/a_2 + 2(k_1/k_2)(d_1d_2/a_1)(1 + k_2/\eta d_2)], \qquad (9)$$

where $\eta$ is the interface thermal conductivity ($k_{1,2}/d_{1,2} \leq \eta < \infty$).

It follows from (9) that, for a thermally thin system, $a_T^{eff}$ depends on $\eta$ and on $k_1/k_2$ in an inverse way and does not depends on the modulation frequency $\omega$.

For $\alpha_T^{eff}$ of a thermally thick two-layer system, a more complex expression has been derived [125]:

$$\alpha_T^{eff} = \omega d^2/2\{arctan[\varphi_+(d_{1,2}, k_{1,2}, \lambda_{T1,2}, \eta)/(\varphi_-(d_{1,2}, k_{1,2}, \lambda_{T1,2}, \eta)] \times [tan(d_1/\lambda_{T1} + d_2/\lambda_{T2})]\}^2, \qquad (10)$$

where $\varphi_+(d_{1,2}, k_{1,2}, \lambda_{T1,2}, \eta) = 1 + k_1\lambda_{T2}/k_2\lambda_{T1} + (k_1/2\eta\lambda_{T1})[1 + ctan(d_1/\lambda_{T1} + d_2/\lambda_{T2})]$ and

$\varphi_-(d_{1,2}, k_{1,2}, \lambda_{T1,2}, \eta) = 1 + k_1\lambda_{T2}/k_2\lambda_{T1} + (k_1/2\eta\lambda_{T1})[1 - tan(d_1/\lambda_{T1} + d_2/\lambda_{T2})]$ are both $\omega$-dependent, since $\lambda_{T1,2} = (2a_{T1,2}/\omega)^{1/2}$.

Thus, in accordance with [125], for a thermally thick two-layer system, $\alpha_T^{eff}$ depends on $\eta$ and on $\omega$. At $\eta \rightarrow \infty$ (ideal interface thermal contact), the dependence (10) transforms to the expression $d/\sqrt{a_T^{eff}} = d_1/\sqrt{a_1} + d_2/\sqrt{a_2}$ proposed in the Ref. [124] and $\alpha_T^{eff}$ is $\omega$-independent. At finite $\eta$ (non-ideal interface thermal contact), the dependence of $\alpha_T^{eff}$ on $\omega$ given by (10) is qualitatively similar to the dependence given by (5 - 7) represented in the form (8).

Thus, the effective thermal diffusivity of two-layer in series systems depends in a complex way on relative thickness fraction of materials of layers, their thermal parameters, the interface thermal conductivity and on the modulation frequency.

It is obvious that the case of TEG-CNT-cs, due to high degree of porosity (**Table 1**) and due to existence of air micro-pockets (**Fig. 1**) is more complicated than the case of two-layer systems considered in the Refs [119, 121 - 125]. Therefore, TEG-CNT-cs can only approximately be represented by a combination of multiple cells of in series and in parallel two-layer systems. At the same time, for TEG-CNT-cs, the measured effective $\alpha_T^{eff}$ values must depend on the geometry and thermophysical parameters of EG-plates and CNTs, as well as on those of the contact layers that gives finite $\eta$ values in [119, 125].

Estimation of the $a_T^{eff}$ values according to (5 - 7) for two-layer systems such as those considered in Refs [121, 123], with layer components similar to TEG-CNT-cs (**Tables 2, 3**), shows that the $a_T^{eff}$ values measured at $f = 10$ Hz and at $f = 20$ Hz can differ from 1.5 to 10 times, depending on the configuration of the two-layer system (Air - TEG or TEG - Air) and on the specific values of thermal conductivity, diffusivity and effusivity of the TEG layers.

In particular, under $a_T^{eff}$ estimation according to (5) at $f = \omega/2\pi$ change from 10 Hz to 20 Hz and using the data obtained by both HPM and TWM (**Tables 2, 3**), $a_T^{eff}$ decreases in ≈1.5 - 10 times for Air - TEG configuration and $a_T^{eff}$ increases in ≈1.5 - 3 times for TEG - Air configuration.

However, under $a_T^{eff}$ estimation according to (7), for both Air - TEG and TEG - Air configurations $a_T^{eff}$ increases in ≈ 1.5 time if the data obtained by HPM are used, and $a_T^{eff}$ decreases in ≈2 - 10 times if the data obtained by TWM are used. This is associated with the different sign of logarithmic term in (7) owing to difference in $b_T$ values obtained by HPM and TWM (see **Tables 2, 3** and **Figs 4, 5**).

Thus, for TEG-CNT-cs, the $\alpha_T^{eff}(\omega)$ dependences similar to (5 - 7) and (10), can be assumed to explain the different $\alpha_T^{eff}$ values obtained by the HPM at (100 ms) and the TWM at (20 Hz) (see **Tables 2, 3**). Therefore, as for two-layer systems [119], the $\alpha_T^{eff}$ values for TEG-CNT-cs may not be unambiguous in terms of their frequency dependence.

Comparison of the data in **Tables 2, 3** shows that the $\alpha_T^{eff}$ values obtained by the HPM [$(10^{-6} - 10^{-5})$ m$^2$/s] are integrally lower and the $\alpha_T^{eff}$ values obtained by the TWM [$(10^{-5} - 10^{-4})$ m$^2$/s] are integrally higher than the $a_T$ values of the air ($2.2 \cdot 10^{-5}$ m$^2$/s). It can therefore be assumed that, at least in the case of TEG-CNT-cs, the HPM is more sensitive to the slow (air) heat transfer component with a low $a_T$ value, while the TWM is more sensitive to the fast (carbon) heat transfer component with a high $a_T$ value.

**CONCLUSION**

The thermally exfoliated graphite (TEG) - multiwall carbon nanotubes (CNT) composites (TEG-CNT-cs) were synthesized by persulphate oxidation method. The processes of deagglomeration of CNT and intercalation of natural graphite were united during electrochemical (anode) oxidation (ECAO) and chemical oxidation (CO). The thin plate samples of TEG-CNT were manufactured by rolling of the powder.

TEM observation of TEG-CNT composites shows the multi-layered structure of partially overlapped single layer graphene nanosheets with folded, wrinkled and tubular-like fragments.

For estimation the effective diffusivity $a_T$ values, the heat pulse (HPM) (100 ms) and thermo-wave modulation (TWM) (20 Hz) methods were applied. The light emitting diode as radiation heater of front surface and the pyroelectric detector as heat sensor of rear surface of TEG-CNT composite sample were used.

For different samples of TEG-CNT-cs, obtained $a_T$ values are ~ $(10^{-6} - 10^{-4})$ m$^2$/s, and estimated $b_T$ values are ~ $(10^3 - 10^4)$ Ws$^{1/2}$/m$^2$K and $k_T$ values are ~ $(1 - 100)$ W/m.K.

The $a_T$ values are comparable with $a_T$ values known for multi-walled CNT sheets in the directions perpendicular and parallel to the CNT alignment in the sheet [62, 63], as well as with $a_T$ values known for graphene multilayers [64] and nanoplatelets [65].

The $b_T$ values are in the ranges characteristic for SiO$_2$, Si, GaAs, Al, Cu, as well as for polymer/graphene nanocomposites [100, 101].

The $k_T$ values are in the range known for graphene/thermoplastic polymer composites [30] and for in-plane and through-plane thermal conductivity of C-C-composites [26].

Not very high $a_T$-, $b_T$-and $k_T$- values of TEG-CNT-cs in comparison with those of regular graphene derivative materials [20, 48] are apparently associated with a weak thermal coupling between the irregularly stacked graphene species in the EG flakes of TEG-CNT composites playing role of a "bottleneck" for heat propagation [102].

For TEG-CNT-cs, evaluated values of phonon mean free path $l_p \approx (0.5 - 40)$ nm and relaxation time $\tau_p \approx (0.05 - 4)$ ps are in the ranges estimated for phonon scattering in defective graphene layers [110]. It is considered as a consequence of far from ideal stacking and state of EG platelets giving rice to high boundary, edge and interlayer phonon scattering.

The difference of $a_T$ values obtained by HPM and by TWM is considered as associated with the difference in the manner of sample excitation in these methods.

Under HPM conditions, when the temperature wave packet spreads, the value of $a_T$ can be underestimated due to the finite pulse time effect [90 - 92].

Under the conditions of TWM, during the propagation of a sinusoidal temperature wave, for TEG-CNT-cs, considered as a set of several two-layer systems, the values of $a_T$ may not be unambiguous in their frequency dependence, as for real two-layer systems [120].

It is found that the values of $a_T$, $b_T$, $l_p$ and $\tau_p$ are lower for denser TEG-CNT-cs samples with high G-fraction obtained by ECAO and are higher for the less dense TEG-CNT-cs samples with low G-fraction obtained by CO. This is associated with the difference of the amount of intrinsic, interface and edge defects in TEG obtained by ECAO and obtained by CO.

Thus, it has been shown that for TEG-CNT-cs, as well as for other C-C-composites [48], the efficiency of their operation as thermal interface and/or heat sink materials depends not only on the thermal properties of their components (EG flakes and CNTs), but also on the thermal quality of their interfaces, which is determined by the TEG-CNT-x preparation method, in particular, the ECAO processing with electrical treatment or the CO processing without it.